\documentclass{article}
\usepackage{graphicx} 
\usepackage[colorlinks]{hyperref}
\usepackage[natbibapa]{apacite}
\usepackage[margin=3.2cm]{geometry}
\usepackage{caption}
\hypersetup{
  citecolor=black,
  urlcolor=blue
}
\title{To democratize research with sensitive data, we should make synthetic data more accessible}
\author{Erik-Jan van Kesteren}
\date{March 2024}

\begin{document}

\maketitle

\begin{abstract}
    For over 30 years, synthetic data has been heralded as a promising solution to make sensitive datasets accessible. However, despite much research effort and several high-profile use-cases, the widespread adoption of synthetic data as a tool for open, accessible, reproducible research with sensitive data is still a distant dream. In this opinion, Erik-Jan van Kesteren, head of the \href{https://odissei-soda.nl}{ODISSEI Social Data Science team}, argues that in order to progress towards widespread adoption of synthetic data as a privacy enhancing technology, the data science research community should shift focus away from developing better synthesis methods: instead, it should develop accessible tools, educate peers, and publish small-scale case studies.
\end{abstract}

Over 30 years ago, \citet{rubin1993statistical} proposed synthetic data as a promising solution to make sensitive datasets available for reuse. The idea is as follows: instead of sharing a sensitive dataset, make available a synthetic version which has similar structure and statistical characteristics. This version can be used for research and analysis, but does not disclose private information of the participants. Indeed, increasing availability of sensitive data holds great promise: a wealth of information is hiding behind access barriers and privacy regulations. If all of this information could be made accessible through public synthetic datasets, this would be an enormous boon for a wide range of scientific disciplines -- especially those involving human subjects (e.g., social, behavioral, medical sciences). 

In my role as the head of the data science team of the Dutch social science data infrastructure (\href{https://odissei-data.nl}{ODISSEI}), I frequently encounter the promise of accessible sensitive data. I consult social scientists on a wide range of projects that would benefit greatly from reduced access barriers. For example, scientists need sensitive data access to (a) analyze historical microdata on residence, demographics, and occupation to research how inequality is distributed geographically, (b) combine multiple existing longitudinal studies to determine lasting psychological impacts of the COVID-19 pandemic, or (c) track school- and student-level metrics from educational registries to assess the impact of an educational policy. Impactful research often needs access to existing sensitive data -- which is costly and time-consuming -- so the promise of synthetic data to ease or even bypass access barriers captivates many social scientists.

Yet, this promise has not led to widespread uptake of synthetic data as a privacy-enhancing tool for improving the accessibility of sensitive data, despite much research effort in the statistics, machine learning, and computer science communities \citep[for recent reviews, see][]{ragunathan2021synthetic, jordon2022synthetic, reiter2023synthetic, fonseca2023tabular, drechsler202330, hu2024advancing}. High-profile cases do exist, such as the \href{https://iknl.nl/nkr/cijfers-op-maat/synthetische-dataset}{Netherlands Cancer Registry}, the \href{https://doi.org/10.11581/yk6n-b652}{The CRPD Cardiovascular disease dataset}, and \href{https://ec.europa.eu/eurostat/web/microdata/public-microdata/statistics-on-income-and-living-conditions}{Eurostat's statistics on income and living public use files}, but most of these have been large projects enacted by teams of data scientists and privacy experts -- not by smaller research groups or individual researchers collecting sensitive data.

The main underlying cause for this is that it is complex, if not impossible, to create a privacy-safe synthetic version of a sensitive dataset that can be both openly publicized \textit{and} productively used for research. The fundamental constraint is the \textit{privacy-fidelity trade-off}: synthetic data is not inherently private \citep{jordon2022synthetic}, and increasing the privacy level requires reducing the statistical closeness (``fidelity'') of the synthetic data to the real data. 

\begin{figure}
    \centering
    \includegraphics[width=0.5\linewidth]{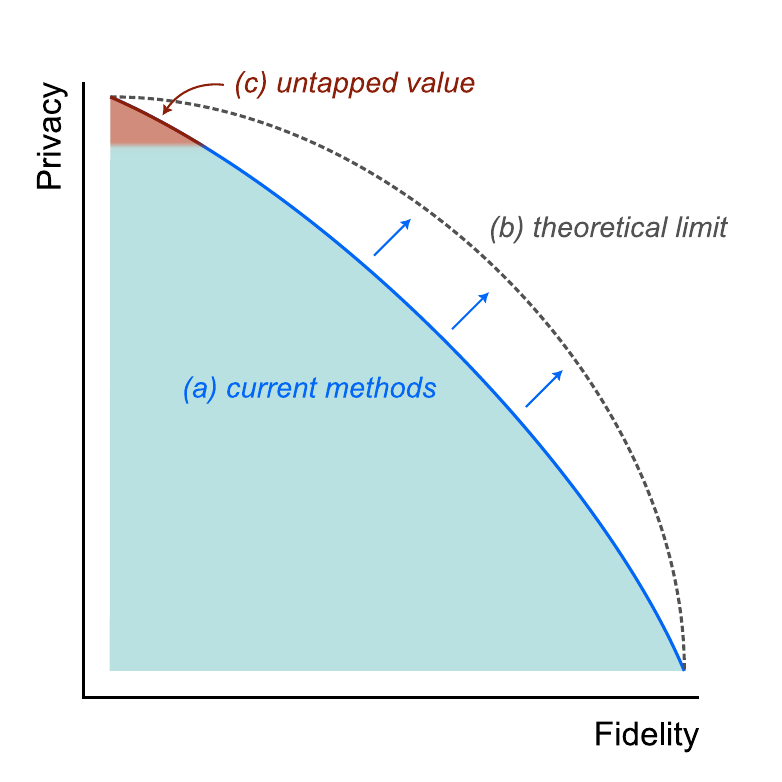}
    \caption{The inverse relation between privacy and fidelity in synthetic data. Several research lines are (a) pushing the currently available methods frontier closer to the theoretical boundary of this privacy-fidelity trade-off, and (b) defining how to measure privacy and fidelity and how to define this theoretical boundary, but they leave a large source of value untapped: (c) low-fidelity synthetic data, which can already be used as a privacy-enhancing technology.}
    \label{fig:frontier}
\end{figure}

A straightforward response is to suggest that we need to research better methods to create high-fidelity synthetic data that adheres to certain privacy standards. Several excellent tools exist which attempt this for a wide variety of underlying generative models and privacy definitions \citep[e.g.][]{nowok2016synthpop, patki2016synthetic, qian2023synthcity} -- their development and evaluation in terms of privacy disclosure risks are the subject of several active research lines. However, while this work is undeniably worthwhile, the best we can hope for with these tools is that they push the privacy-fidelity frontier towards its theoretical limit (Figure \ref{fig:frontier}). As such, precluding a major revolution in our statistical definition and treatment of privacy, these advances are not bringing us closer to the widespread accessibility of sensitive data in a privacy-friendly way.

To put it more concretely, the current lines of data science research in synthetic data are focused too heavily on the technical aspects of synthetic data generation \citep{hu2024advancing}. Putting synthetic data into real-world practice requires solving real-world issues, such as legal compliance, organizational reputation\footnotemark, education of data managers and owners, infrastructure development, training researchers to use this data, and more.

\footnotetext{Anecdotally, a national statistical agency publishing synthetic version of their microdata might be concerned with the question \textit{``what if the public thinks this is real data and publishes analyses using this data?''}}

\,

To start tackling these issues, I suggest a renewed focus from the data science community towards the safe side of the trade-off: create low-fidelity data with high privacy\footnotemark. I consciously use the word \textit{fidelity} instead of the more commonly used term \textit{utility}, because I think low-fidelity synthetic data is of very high utility, for example in research code development, education, open science, reproducible research, and data accessibility. One of the biggest problems with research on sensitive data is that it takes time, effort, and money for researchers to be allowed access. Synthetic data can reduce these costs. With a public, low-fidelity mock version of a sensitive dataset, it is possible to inspect its columns, assess whether a certain research question can be answered, and even create analysis scripts that can run on the real data. In this way, the barriers to the first steps in the research process are greatly diminished, thereby democratizing research with existing sensitive data.

\footnotetext{The UK's Office for National Statistics would call such data \textit{synthetically-augmented plausible} in their categorization \citep{bates2019ons}}

Potential ways in which low-fidelity synthetic data can democratize and improve research with sensitive data are as follows: 
\begin{itemize}
    \item Findability and accessibility. An organisation that produces sensitive data may release publicly a low-fidelity synthetic version of the data. This lowers the barrier for researchers to discover and use sensitive data, as it becomes easier to look up and test its contents and structure.
    \item Computational reproducibility. Upon publication of a project which uses sensitive data, researchers may publish a low-fidelity synthetic version of the data together with their public code repository (or point to where such data is available). This allows others to run, check, and learn from the code, even though the analysis results will be quite different from those in the original published paper.
    \item Security. If low-fidelity synthetic data is publicly available, researchers can assess exactly which parts they need before asking for access to the real data. With this information, a data owner can restrict subsequent access to only those columns and rows in specific tables. An even stronger security model would be ``code-to-data'', which means running an analysis script on the sensitive data without researcher access. In this approach, low-fidelity synthetic data is essential to developing analysis scripts. These newly accessible security methods may convince data owners to enable researcher access where this was not possible before.
    \item Transparency and rigor. When analysis scripts are developed openly on a low-fidelity synthetic version of the data and then run only once on real data, it is harder for analysts to engage in common questionable research practices which may invalidate scientific findings. Common examples are manually removing data points or selecting variables to make results in line with hypotheses.
\end{itemize}

A perceived downside of low-fidelity synthetic data is that it still requires access to the real, sensitive data in order to be able to make valid conclusions. Indeed, the approach I am advocating renounces one of the most attractive promises of synthetic data, namely the ability to do analyses without needing real data access at all. At first sight, this appears to be a steep price to pay, but I argue that this promise cannot be fulfilled \textit{in general}\footnotemark: it would require the synthetic data to take into account any analysis, including nonlinear, multilevel, and outlier-dependent methods, all of which are hard (or impossible) to safely enable. For sensitive data providers, the goal of allowing access is often to enable valuable new research. In most cases, it is impossible know in advance which methods or statistics can answer valuable research questions, so access to the real data remains needed regardless of the fidelity of the synthetic data. In these cases, low-fidelity synthetic data is a great solution, when combined with other existing privacy enhancing infrastructure such as secure remote computing environments or systems that bring analysis code to the data.

\footnotetext{For certain analysis types this promise can be fulfilled. For example, for many linear models it is sufficient for the synthetic data to accurately represent the means, covariance matrix, and sample size of the sensitive data, which are relatively easy to protect for public release. This can be done by (a) computing these statistics from the sensitive data, (b) performing an appropriate level of sufficient statistics perturbation (SSP) or disclosure control, and (c) generating multivariate normal synthetic data based on the perturbed statistics.}

\begin{figure}[h]
    \centering
    \includegraphics[width=0.7\linewidth]{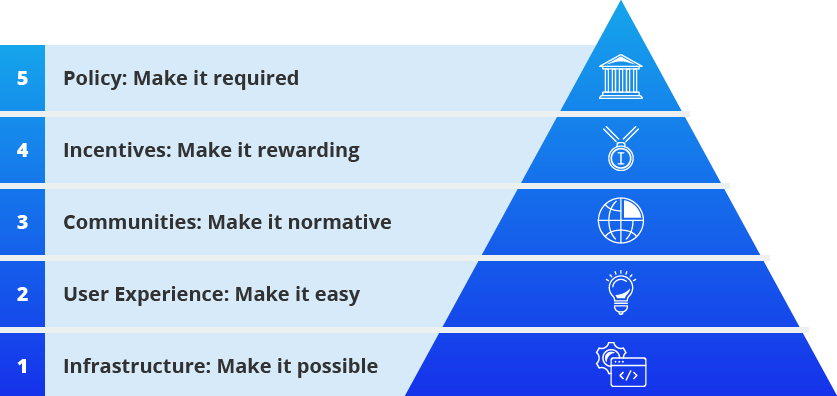}
    \caption{Steps to strategic change in science. Reprinted from \url{https://www.cos.io/blog/cos-celebrates-10-years} (accessed 2024-07-03).}
    \label{fig:cos-change}
\end{figure}

To capitalize on the value of synthetic data as a privacy enhancing technology, we as a data science community should work on including it in real-world scientific practice. My proposal of focusing more on accessibility aligns with the theory of strategic change in science put forth by the center for open science (see Figure \ref{fig:cos-change}). The past 30 years in synthetic data research have focused on ``making it possible'' (step 1); in order to extract the value of synthetic data and democratize research with sensitive data it is now time to ``make it easy'' (step 2).

In conclusion, in addition to continuing existing research lines on the privacy-fidelity trade-off for synthetic data, the data science community should work directly on removing the barriers for data owners to create an open synthetic version of their data for public use. The community should build accessible tools \citep{jordon2022synthetic}, develop educational materials \citep{hu2024advancing}, and implement and publish pilot projects. Promising projects in this regard do already exist: auditable and accessible synthetic data tools such as \href{https://github.com/sodascience/metasyn}{\textit{metasyn}} \citep{schram_2024_10869586}, educational projects such as the Urban Institute's trainings and associated reports \citep{pickens2023generating}, and use-cases of synthetic data for computational reproducibility can be found on the Open Science Framework \citep[e.g.,][]{Kirtley_Hussey_Marzano_2023}. Let us expand these efforts in order to pave the way for more widespread use of synthetic data for open, accessible, reproducible research with sensitive datasets.

\section*{Acknowledgments}
I am grateful for valuable feedback from Peter-Paul de Wolf, Thom Benjamin Volker, and Lara Rösler.

\bibliography{biblio}

\end{document}